\documentclass[preprintnumbers,twocolumn,nofootinbib,amsmath,amssymb,aps,nofootinbib]{revtex4-1} 
\usepackage{graphicx}
\usepackage{dcolumn}
\usepackage[utf8]{inputenc}
\usepackage{bm}
\usepackage[caption=false]{subfig}
\usepackage{amssymb}
\usepackage{float}
\usepackage{hyperref}
\usepackage[T1]{fontenc} 
\usepackage{subfig}
\usepackage{dsfont}
\usepackage{slashed}
\usepackage{color}
\usepackage{amsmath}
\usepackage{mathtools}
\usepackage[section]{placeins}
\usepackage{braket}
\usepackage{upgreek}
\usepackage[bottom]{footmisc}
\usepackage[normalem]{ulem}
\usepackage{lipsum}
\usepackage[utf8]{inputenc}

\newcommand{\tr}{\mathrm{tr}}

\def\tr{{\rm Tr}}

\begin{document}

\title{Quantum simulation of Markovian and non-Markovian open quantum dynamics in heavy-ion collisions} 

\author{Doojin Kim}
\email{doojin.kim@usd.edu}
\affiliation{Department of Physics, University of South Dakota, Vermillion, SD 57069, USA}

\author{Balbeer Singh}
\email{balbeers@lip.pt}
\affiliation{LIP, Av. Prof. Gama Pinto, 2, P-1649-003 Lisboa, Portugal }

\begin{abstract}
We present a quantum computing framework for simulating open-quantum-system approaches based on Markovian and non-Markovian dynamics, which is relevant to heavy-ion collisions.  To simulate the non-Markovian evolution on quantum computers, we introduce an auxiliary two-level pseudomode that carries the memory forward and couples to both the subsystem and the residual Markovian bath. We explicitly show that tracing out the pseudomode reproduces the non-Markovian evolution with the exact memory kernel. Moreover, in the relevant time scale hierarchy, the quantum circuit construction of the pseudomode smoothly converges to the Markovian limit. For a given bath memory kernel, our results demonstrate the feasibility of quantum simulations of both Markovian and non-Markovian dynamics, establishing a framework for future studies of hard probes such as jets, heavy quarks, and quarkonia in heavy-ion collisions.
\end{abstract}

\maketitle

\section{\label{sec:introduction}Introduction}
Quantum computing has emerged as a promising framework for simulating quantum dynamics that can become increasingly demanding for classical computational methods, particularly when real-time evolution, many-body correlations, and coupling to environmental degrees of freedom must be treated explicitly~\cite{Bauer:2022hpo,Preskill:2018fag,Guo:2026nuc}. An important class of such problems is provided by open quantum systems (OQSes), for which quantum algorithms offer a natural route to representing unitary system-environment dynamics and the resulting non-unitary reduced evolution of a subsystem~\cite{Delgado-Granados:2024rab}. These ideas have recently attracted growing interest in nuclear and high-energy physics, including applications to the real-time dynamics of hard probes in relativistic heavy-ion collisions (HICs)~\cite{deJong2021,Chen:2026tvd,Yao:2025uxz,Lee:2024jnt,Wei:2026iwq,deJong:2021wsd,Castro:2025ocx,Barata:2023clv}. This motivates the development of quantum-simulation frameworks that go beyond the commonly studied Markovian limit and can retain finite environmental memory in a controlled and circuit-compatible manner~\cite{Breuer2016,Li:2021ssn,Li:2024kkr}.

OQSes are coupled systems in which a subsystem interacts with an environment, or bath, typically possessing infinitely many degrees of freedom~\cite{BreuerPetruccioneBook,Gorini1976}. While the combined subsystem-bath evolution is generally intractable to solve,  one instead traces out the bath degrees of freedom to obtain a reduced description of the subsystem. In the Markovian limit when the bath correlation time $\tau_E$ is small compared to the subsystem time $\tau_S$ scale, i.e., $\tau_E\ll \tau_S$, the reduced evolution is described by the Lindblad equation~\cite{Lindblad1976}. On the other hand, when the bath correlation time is comparable to the subsystem time scale, memory effects become important, leading to non-Markovian evolution of the reduced density matrix~\cite{Breuer2016}. The open system dynamics and the interplay between Markovian and non-Markovian regimes arise across a wide range of physical systems from cosmology, neutrino physics to HICs~\cite{Hollowood:2017bil,Stankevich:2024xyc,Akamatsu:2020ypb}.

High-energy nuclear collision experiments create a short-lived, deconfined state of quarks and gluons known as quark-gluon plasma (QGP) at temperatures as large as $T\sim450$~MeV~\cite{PHENIXWhitePaper2005,STARWhitePaper2005}.  The properties of QGP are inferred indirectly through probes that traverse the medium and retain imprints of their interaction with the plasma~\cite{Apolinario:2022vzg}. The OQS framework has become a widely used tool for hard probes of QGP as it naturally separates the subsystem of interest from the surrounding medium and describes how the reduced dynamics of the subsystem is modified by the presence of QGP. The OQS approach has been applied extensively to quarkonium dynamics~\cite{Armesto:2026fit,Akamatsu2012,Brambilla2018,Yao2019,Yang:2024ejk,Delorme:2024rdo,Strickland:2023nfm,Yao:2021lus} as well as transport properties of heavy quarks~\cite{Akamatsu2015}. More recently, the OQS framework has also been expanded to study jet production and evolution in HICs~\cite{Yao2021,Mehtar-Tani:2024smp,Mehtar-Tani:2025xxd,Arleo:2026urv}. For more applications of OQS, see Refs.~\cite{Barata:2026hck,BlaizotEscobedo2018a,BlaizotEscobedo2018b,Armesto:2025usb,Lee:2023urk,Barata:2023uoi}.  While the applications of the OQS framework in HICs are growing, we also need computational methods that allow us to solve evolution equations as precisely as possible.

In this paper, we simulate OQS dynamics relevant to HICs on quantum computers. To this end, we work in a simplified setup of a two-level bound state interacting with a static and homogeneous thermal medium. Within this setup, we simulate both Markovian and non-Markovian evolution equations. While the Markovian regime has been extensively studied in the literature, the non-Markovian regime remains comparatively unexplored. For quarkonium, it has been argued that non-Markovian effects become important during the late stages of the collision when the temperature of the plasma drops down due to hydrodynamic expansion, and the separation of timescales leading to the Markovian approximation breaks down. Moreover, non-Markovian dynamics is also important in the diffusion of heavy quarks as well as in the evolution of extended multi-partonic objects such as jets in QGP~\cite{BR:2025lhx,Sharma:2023dhj,Singh:2024pwr,Nair:2026tqy,Sharma:2025vvi}. 

To implement the non-Markovian evolution on a quantum circuit, we consider the medium correlator to decay exponentially in time and introduce an auxiliary two-level mode, or a pseudomode~\cite{Dalton2001,Mazzola2009,Pleasance2020,Imamoglu1994,Garraway1997a,Garraway1997b,Tamascelli2018}. The pseudomode is constructed such that its relaxation rate is the same as the decay rate of the memory kernel. The pseudomode is coupled to the subsystem, i.e., the bound state, through a coherent interaction that exchanges excitations between the two, and separately coupled to a Markovian bath through a dissipator that models loss of coherence of the pseudomode. As a result, the pseudomode evolves jointly with the subsystem and is carried forward from one time step to the next, retaining the memory, whereas the bath coupled to the pseudomode is discarded after every time step and reinitialized before the next. We explicitly show that in the relevant time scale hierarchies discussed above, the non-Markovian pseudomode construction reduces to its Markovian limit.   

The rest of the paper is organized as follows. In section~\ref{sec:setup}, we briefly discuss the OQS framework and derive evolution equations for both Markovian and non-Markovian dynamics within simplified assumptions for a bound state. In section~\ref{sec:markovian} we discuss the simulation of the Lindblad equation with an exponentially decaying medium correlator on quantum circuits and compare the survival probability of the bound state with RK4 solutions. In section~\ref{sec:nonmarkovian} we discuss pseudomode construction for simulating non-Markovian dynamics on quantum circuits and obtain the convergence of pseudomode construction with the memoryless limit. Finally, in section~\ref{sec:summary} we summarize our results and discuss the future applications of quantum computing methods for various hard probes in HICs.

\section{Set-up}
\label{sec:setup}
We consider a particle of mass $M$ and its antiparticle pair, e.g., a heavy positronium-like system interacting with a stationary thermal bath at temperature $T$ characterized by an exponentially decaying medium correlator. It is well known that such massive pairs can be treated within the non-relativistic effective field theories. While the specific details of the theory are not relevant in this paper to derive the analytical equations, we take the potential non-relativistic quantum electrodynamics (pNRQED) framework. For a nonrelativistic bound state, the characteristic scales, the binding energy $E_b$ and the inverse size of the bound state $1/r$, satisfy $E_b \sim Mv^2 \ll 1/r \sim Mv$. We further assume that the thermal scale is below the inverse size, $T\ll 1/r$, so that the interaction with the medium can be organized as a multipole expansion in the relative coordinate $\mathbf{r}$.
At leading order in this expansion, the Lagrangian takes the form~\cite{Brambilla2005pNRQCD} 
\begin{equation}
\mathcal{L} = \mathcal{L}_S + \mathcal{L}_E + \tilde{\mathcal{L}},
\label{eq:lagrangian}
\end{equation}
where $\mathcal{L}_S$ describes the free propagation of the heavy-pair
subsystem, $\mathcal{L}_E$ describes interactions in the bath, and $\tilde{\mathcal{L}}$ encodes the leading-order interaction between the subsystem and the medium.\footnote {Throughout this work, we use ``medium'' to refer to the physical QGP/plasma, while ``bath'' denotes its role as the environment in the OQS description. Unless confusion may arise, we use the two terms interchangeably.} At leading order in the multipole expansion, this is a dipole interaction and reads
\begin{equation}
\tilde{\mathcal{L}} \equiv \left( S^\dagger\, e \mathbf{r}\cdot\mathbf{E}\, O
+ O^\dagger\, e \mathbf{r}\cdot\mathbf{E}\, S \right),
\label{eq:Lint}
\end{equation}
where $e$ is the electromagnetic coupling constant, $\mathbf E$ is the electric field sourced by the plasma, and $e\mathbf r\cdot\mathbf E$ is the standard dipole coupling. The fields $S$ ($S^\dagger$) and $O$ ($O^\dagger$) represent the bound and unbound
(free/continuum) configurations of the heavy quark pairs, taking care of the destruction (creation) of the bound and unbound states, respectively. Physically, Eq.~\eqref{eq:Lint} is the operator responsible for driving transitions between bound and unbound states through the plasma. 
We emphasize that Eqs.~\eqref{eq:lagrangian} and \eqref{eq:Lint} are introduced only for the purpose of deriving the evolution equation to verify the construction of quantum circuits. We leave the incorporation of explicit details of the effective field theory and thermal medium for future work.

Having decomposed the heavy-pair field into bound and continuum components, we now turn to its evolution resulting from the interactions with the plasma. 
To this end, we construct the total Hamiltonian of the system (subsystem+bath) from the Lagrangian as 
\begin{equation}
H = H_S + H_E + \tilde{H} \equiv H_0 + \tilde{H},
\end{equation}
where $H_S$ describes free evolution of the heavy pairs, $H_E$ describes the interaction within the plasma, and $\tilde{H}$ couples the bath and the subsystem which is obtained from $\tilde{\mathcal{L}}$ given in Eq.~\eqref{eq:Lint}. Since we are interested in the effect of the medium on the subsystem rather than the free evolution of either piece separately, it is convenient to move to the interaction picture, where the density
matrix of the full system evolves under $\tilde{H}$
\begin{equation}
\rho_I(t) = U_I(t)\, \rho_I(0)\, U_I^\dagger(t),
\label{eq:rhoi}
\end{equation}
where the subscript $I$ denotes that the operators are dressed with the free Hamiltonian. 
An operator is then transformed as $\mathcal{O}_I(t) = U_0^\dagger(t)\, \mathcal{O}\, U_0(t)$ with $U_0(t)=e^{-iH_0 t}$, e.g., $\tilde{H}_I(t)=U_0^\dagger(t)\, \tilde{H}\, U_0(t)$. The remaining time evolution in the interaction picture is generated by $\tilde{H}_I(t)$ with 
\begin{equation}
U_I(t) = \mathcal{T}\exp\left[-i\int_0^t dt_1\,\tilde{H}_{I}(t_1)\right]    
\end{equation}
where $\mathcal{T}$ denotes time ordering.   Differentiating Eq.~\eqref{eq:rhoi} with respect to time and using the relation $\dot U_I(t)=-i\tilde{H}_{I}(t)U_I(t)$, we get $\dot\rho_I(t) = -i[\tilde{H}_{I}(t),\rho_I(t)]$. Next integrating this once and resubstituting the resulting expression for $\rho_I(t)$ back into $\dot\rho_I(t)$, we get
\begin{equation}
\frac{d\rho_I(t)}{dt} = -i[\tilde{H}_{I}(t), \rho_I(0)]
- \int_0^t ds\, \big[\tilde{H}_{I}(t), [\tilde{H}_{I}(s), \rho_I(s)]\big]. \label{eq:rhoIdt}
\end{equation}
This equation still couples the subsystem to the full plasma density
matrix. To obtain the reduced density matrix of the subsystem, we assume the heavy pair is weakly
coupled to the plasma. We can therefore write the total density matrix in a factorized form as $\rho(t) \simeq \rho_S(t) \otimes \rho_E$ where  $\rho_E$ is the thermal density matrix of the medium and $\rho_S$ is the subsystem density matrix. Writing $\tilde{H}_{I}(t)\equiv\sum_i d_{i}(t) E_{i}(t)$ and tracing out the bath degrees of freedom, we obtain the reduced subsystem density matrix that reads as 
\begin{align}
\frac{d\rho_S}{dt} &= -\sum_{ij} \int_0^t ds\,
\Big[ C_{ij}(t-s)\, [d_i(t),\, d_j(s)\rho_S(s)] \nonumber \\
&\qquad + C_{ji}(s-t)\, [\rho_S(s) d_j(s),\, d_i(t)] \Big],
\label{eq:master}
\end{align}
where $d_i = e r_i$ is the dipole operator and $C_{ij}(t) = {\rm Tr}[E_i(t) E_j(0) \rho_E]$ is the electric-field correlator, which essentially encodes information about the medium relevant to the evolution of the subsystem.
Since the thermal bath is assumed to be stationary, its correlation functions are invariant under time translations and therefore depend only on the time difference, $C_{ij}(t,s)=C_{ij}(t-s)$.
Note that the contribution from the first term in Eq.~\eqref{eq:rhoIdt} disappears as we assume that the thermal bath has vanishing one-point functions, ${\rm Tr}[E_i(t) \rho_E]=0$, so that the first-order contribution vanishes.

We emphasize that Eq.~\eqref{eq:master} retains the full structure of the medium correlator, and that at any given time $t$ the integrand depends on the complete history $\rho_S(s)$ for all $s<t$, rather than on $\rho_S(t)$ alone. In the following sections, we solve this equation in two physically distinct limits: first, the regime in which the bath loses memory of its own fluctuations much faster than the subsystem evolves, which yields a time-local (Markovian) description; and second, the regime in which the bath retains memory over timescales comparable to the subsystem's evolution, so that this memory directly affects the subsystem dynamics and a genuinely non-Markovian treatment is required. For the Markovian approximation, we take the hierarchy $\tau_E \ll \tau_S$, where $\tau_E \sim 1/T$ is the bath correlation time and $\tau_S \sim 1/ E_b$ is the subsystem timescale that is set by the binding energy of the bound state.  Under this approximation, one may (i) replace $\rho_S(s) \to \rho_S(t)$ inside the integral and (ii) extend the upper limit of integration in Eq.~\eqref{eq:master} to infinity.  

Writing the dipole operator in terms of the bound and unbound projection operators, redefining $\tau=t-s$, and dropping terms that oscillate rapidly in $t$, we obtain a time-local Lindblad form of the evolution equation as~\cite{Akamatsu:2020ypb}
\begin{equation}
\frac{d\rho_S}{dt}
= -i\left[H_{\rm eff}, \rho_S\right]
+ \sum_{ij} \gamma_{ij}(\omega)
\left( L_j \rho_S L_i^\dagger
- \frac{1}{2}\left\{ L_i^\dagger L_j, \rho_S \right\} \right),
\label{eq:lindblad}
\end{equation}
where $H_{\rm eff} = H_S + H_{\rm LS}$, with $H_{\rm LS}$ being the correction to the energy of the subsystem which we assume to be negligible for illustration and $\gamma_{ij}$ is the real part of Fourier transfrom of the medium correlator $C_{ij}$. The operators $L_i, L_j \propto |f\rangle\langle b|$, where $|b\rangle$ denotes a bound state and $|f \rangle$ unbound states, are jump operators. The quantity $\omega$ is governed by the energy difference $\Delta E$ between the bound and unbound states. As mentioned earlier, the hierarchy, $\tau_E\ll \tau_S$, leads to a Markovian approximation. By contrast, the Lindblad equation of the reduced density matrix breaks down when $E_b\sim T$. In this case, we must retain the full non-Markovian structure of Eq.~\eqref{eq:master}. In realistic scenarios, this could be realized in the late stages of HICs where the QGP eventually cools down, leading to $E_b\sim T$. Without any loss of generality, we choose an exponentially decaying profile of the medium correlator given as\footnote{This form of memory kernel has been used for investigating non-Markovian effects on heavy quark transport properties~\cite{Nair:2026tqy}.}
\begin{equation}
C(\tau)=C_0e^{-|\tau|/\tau_c},
\label{eq:mcorr}
\end{equation}
where $C_0=C(0)$ sets the overall strength of the medium correlation function, while $\tau_c$ determines its temporal decay and hence the memory time of the medium.

\section{Quantum computing}
\label{sec:quantum-computing}

In this section, we discuss quantum circuits and a quantum-computing strategy for solving reduced density matrix evolution equations in both the Markovian and non-Markovian regimes. For this, we assign one qubit to the subsystem and use ancilla qubits prepared in a fixed reference state for the bath. The dissipative evolution is then implemented as a unitary circuit via Stinespring dilation with ancilla traced out at the end of each time step to recover the physical evolution of the subsystem. The two regimes (Markovian and non-Markovian) differ mainly in how the resulting generator is treated. In the Markovian case, we get a single, time-independent generator that can be implemented at every step; on the other hand, the non-Markovian case requires an auxiliary pseudomode to encode the memory effects in the bath. To simulate the evolution equations on quantum circuits, we construct Kraus operators obtained by exponentiating the generator, which makes the resulting channel completely positive and trace-preserving for any time step. 

\subsection{Markovian limit}
\label{sec:markovian}

\subsubsection{Markivian open-system evolution}
We first work in the Markovian limit and simulate the reduced density-matrix evolution given by Eq.~\eqref{eq:lindblad}. To simplify the computation and illustrate qubit-based calculation procedures, we restrict to only two levels, i.e., a single bound state and an unbound state. In realistic scenarios, for example of quarkonium, one would need to include additional bound states with different principal and angular quantum numbers, as well as a continuum of unbound states.\footnote{For recent progress on quantum computing for quarkonium states in the Markovian limit, see Refs.~\cite{Wei:2026iwq,deJong2021}.} The application of quantum computing for such multi-state systems is beyond the scope of this study, and we reserve them for future work. 

As mentioned earlier, the dipole operator $d_i$ connects only
$|b\rangle$ and $|f\rangle$ so that we can write 
\begin{equation}
    d_i(t) = d_i^{bf}\,e^{-i\Delta E\,t}\,\hat L^\downarrow + d_i^{fb}\,e^{i\Delta E\,t}\,\hat L^\uparrow,
\end{equation}
where the two jump operators are explicitly given as\footnote{We introduce arrowed symbols to explicitly specify the physical role of the operators; $\hat{L}^\uparrow$ and $\hat{L}^\downarrow$ are identified as $L$ and $L^\dagger$, respectively.}
\begin{equation}
\hat L^\downarrow\equiv|b\rangle\langle f|,\ \qquad \hat L^\uparrow\equiv|f\rangle\langle b|.
\end{equation}
Throughout this work, the transition matrix $d_{i}^{bf}=e\langle b|r_i|f\rangle = \left(d_i^{fb} \right)^*$ is treated as a free parameter rather than evaluated from first principles, which requires performing the overlap integral of the bound-state Coulomb wavefunction with a continuum scattering wavefunction at the relevant relative momentum. We further assume the medium correlator to be isotropic so that the indices $i,j(=x,y,z)$ collapse into a scalar proportional to $\delta_{ij}$, i.e., only diagonal $C_{ij}$'s are non-vanishing:
\begin{equation}
\gamma_{ij}(\omega)\equiv \delta_{ij}\gamma(\omega) =2\,{\rm Re}\Big[\int_0^{\infty} d\tau\, e^{i\omega \tau}\,C_{ij}(\tau)\Big].
\end{equation}
Using this simplification, we obtain 
\begin{equation}
\frac{d\rho_S}{dt} = -i[H_S,\rho_S] + \Gamma_{\rm diss}\,\mathcal D[\hat L^\uparrow]\rho_S + \Gamma_{\rm rec}\,\mathcal D[\hat L^\downarrow]\rho_S,
\label{eq:markovian}
\end{equation}
where the dissipator $\mathcal D[L]\rho = L\rho L^\dagger - \tfrac12\{L^\dagger L,\rho\}$ satisfying the relation $\tr[\mathcal D[L]\rho]=0$. 
The generator is built from the system Hamiltonian $H_S$ and jump operators $\hat L^\downarrow,\hat L^\uparrow$  which can be exponentiated together to obtain the circuit over any time step $\Delta t$. 
The Fourier transform of the correlator given in Eq.~\eqref{eq:mcorr} is given as
\begin{equation}
\gamma(\omega) = \frac{2C_0\,\tau_c^{-1}}{\tau_c^{-2}+\omega^2},
\label{eq:gamma}
\end{equation}
which is evaluated at $\omega=\pm \Delta E$, for $\omega \ll \tau_c^{-1}$ we get $\gamma\to 2C_0\tau_c$. For illustration, we assume $E_f=0$ for an unbound state taken near the continuum threshold, so we identify 
\begin{equation}
    \Delta E=0-(-E_b)=E_b,
\end{equation}
unless stated otherwise, throughout this work.
As mentioned earlier, $\tau_c$ is the time over which the correlator $C(\tau)$ decays, and the medium effectively does not retain the memory. We stress that the Markovian approximation is valid when $\Delta E \tau_c\equiv E_b\tau_c\ll1$. 
Moreover, the rates are given by 
\begin{eqnarray}
    \Gamma_{\rm rec}&=&|d_{bf}|^2\gamma(\Delta E)(1+\bar n), \\
    \Gamma_{\rm diss}&=& |d_{bf}|^2\gamma(\Delta E)\bar n,
\end{eqnarray}
where $\bar{n}$ is the Bose-Einstein distribution function of (photonic) bath constituents and $|d_{bf}|^2\equiv d_{i}^{bf}d_{j}^{bf}\delta_{ij}$. Note that for the simplified real and even correlator adopted here, the spectral function \(\gamma(\omega)\) does not by itself encode thermal detailed balance. We therefore include the standard bosonic occupation factors here. 

\subsubsection{Quantum-circuit implementation}
To solve the Lindblad equation on quantum computers, we exponentiate the generator $\mathcal{O}_{\rm Markov}$, keeping in mind that Eq.~\eqref{eq:markovian} can be understood as an operator equation $\dot{\rho_S}=\mathcal{O}_{\rm Markov}[\rho_S]$. To this end, we first vectorize $\rho_S$ and write it as a single-length four-vector; for example, 
\begin{equation}
    \vec\rho \equiv {\rm vec}\left[
    \begin{pmatrix}
        \rho_{S,00} & \rho_{S,01} \\
        \rho_{S,10} & \rho_{S,11}
    \end{pmatrix}
    \right]=
    \begin{pmatrix}
        \rho_{S,00} \\
        \rho_{S,10} \\
        \rho_{S,01} \\
        \rho_{S,11}
    \end{pmatrix}.
\end{equation}
Using the identity $\mathrm{vec}[AXB]=(B^T\otimes A)\,\mathrm{vec}[X]$ with $\otimes$ symbolizing the Kronecker product, this allows us to rewrite the operator equation as an ordinary linear system, $\dot{\vec\rho}=\mathcal L_{\rm Markov}\vec\rho$, with the explicit $4\times4$ matrix representation for the operator $\mathcal{O}_{\rm Markov}$, i.e., 
\begin{equation}
    \mathcal{L}_{\rm Markov}=-i(I\otimes H-H^T\otimes I)+\Gamma_{\rm diss}\mathcal{D}_{L_{\rm diss}}+\Gamma_{\rm rec}\mathcal{D_{L_{\rm rec}}},
\end{equation}
where $\mathcal{D}_{L_i}$ is defined as
\begin{equation}
    \mathcal{D}_{L_i}=L_i^*\otimes L_i -\frac{1}{2}\left[I\otimes L_i^\dagger L_i + \left(L_i^\dagger L_i\right)^T\otimes I\right],
\end{equation}
with $I$ being the $2\times 2$ identity matrix. The evolution over one time step $\Delta t$ is obtained by matrix exponentiation denoted by $M$,
\begin{equation} 
M(\Delta t) \equiv \exp\big(\mathcal L_{\rm Markov}\Delta t\big),
\end{equation}
leading to the exact solution 
\begin{equation}
    \vec\rho(t+\Delta t)=M(\Delta t)\,\vec\rho(t).
\end{equation}
Since $\mathcal L_{\rm Markov}$ is built from $H_S$ and jump operators with non-negative rates $\Gamma_{\rm diss},\Gamma_{\rm rec}\ge0$, the evolution equation gives a quantum channel that is completely positive and trace-preserving. 

To extract Kraus operators, we rewrite $M(\Delta t)$ into its Choi-Jamiolkowski representation to implement them on a quantum circuit; we use Stinespring dilation and turn the Kraus operators into a single unitary acting on the subsystem together with an ancilla which acts as a bath~\cite{Cleve2017,Wang2011,Sweke2015}.  At the end of each time step, we trace out the ancilla, describing the evolution of the subsystem density matrix governed by Eq.~\eqref{eq:markovian}.

\begin{figure}[h]
\centering 
\includegraphics[width=1\linewidth]{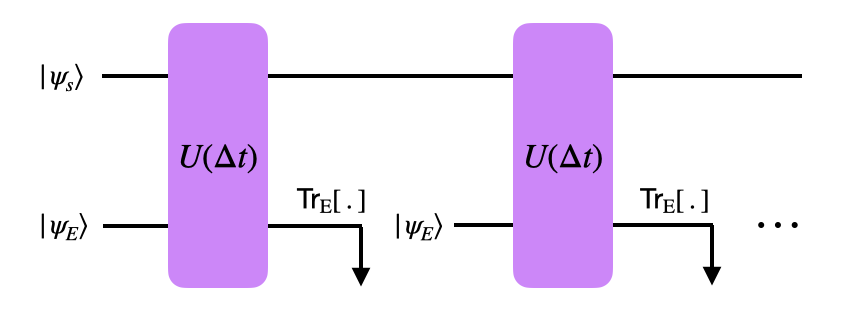}
\caption{A circuit realization of two Markovian time steps. A single unitary $U(\Delta t)$ acts jointly on the subsystem qubit state $|\psi_S\rangle$ and two ancilla qubit states represented by $|\psi_E\rangle\equiv |00\rangle$.
The ancillas are traced out $\mathrm{Tr}_E[\cdot]$ and re-prepared fresh before the next repetition. Iterating this single unitary $N_{\rm step}$ times describes the evolution of Eq.~\eqref{eq:lindblad} over total time $t=N_{\rm step}\Delta t$.} 
\label{fig:markovcircuit}
\end{figure}

From Kraus operators, we first construct a matrix, $V \equiv (K_0;K_1;K_2;K_3)$ that allows us to construct a unitary matrix $U$. The unitary operator then acts on the subsystem together with ancilla qubits, giving 
\begin{equation} 
U(|\psi_S\rangle\otimes|0\rangle_{\rm anc}) = \sum_{k=0}^3(K_k|\psi_S\rangle)\otimes|k\rangle_{\rm anc},
\end{equation}
where $|\psi_S\rangle=\alpha |b\rangle +\beta |f\rangle$ denotes the generic subsystem state and the notation $|k\rangle_{\rm anc}$ collectively denotes states $|00\rangle,|01\rangle,|10\rangle$, and $|11\rangle$. Tracing out ancillas afterward then reproduces the dissipative evolution of the subsystem. In each time step, ancillas are prepared fresh so that they carry no memory between the steps, as schematically shown in Fig.~\ref{fig:markovcircuit}.
Therefore, one Markovian time step is realized as a single unitary $U$ acting on three qubits in total: one qubit for the system itself and two ancilla qubits $\log_2 K=n_{\rm anc}$ where $K$ is the number of Kraus operators.  The unitary is applied once, after which the ancilla qubits are discarded and re-prepared before the next repetition begins.  Because the ancillas carry no information forward from one repetition to the next, each step is completely independent of all earlier ones. Repeating this three-qubit unitary $N_{\rm step}$ times describes the evolution of Eq.~\eqref{eq:lindblad} over the total time $t=N_{\rm step}\Delta t$. 

\begin{figure}[h]
\centering 
\includegraphics[width=1\linewidth]{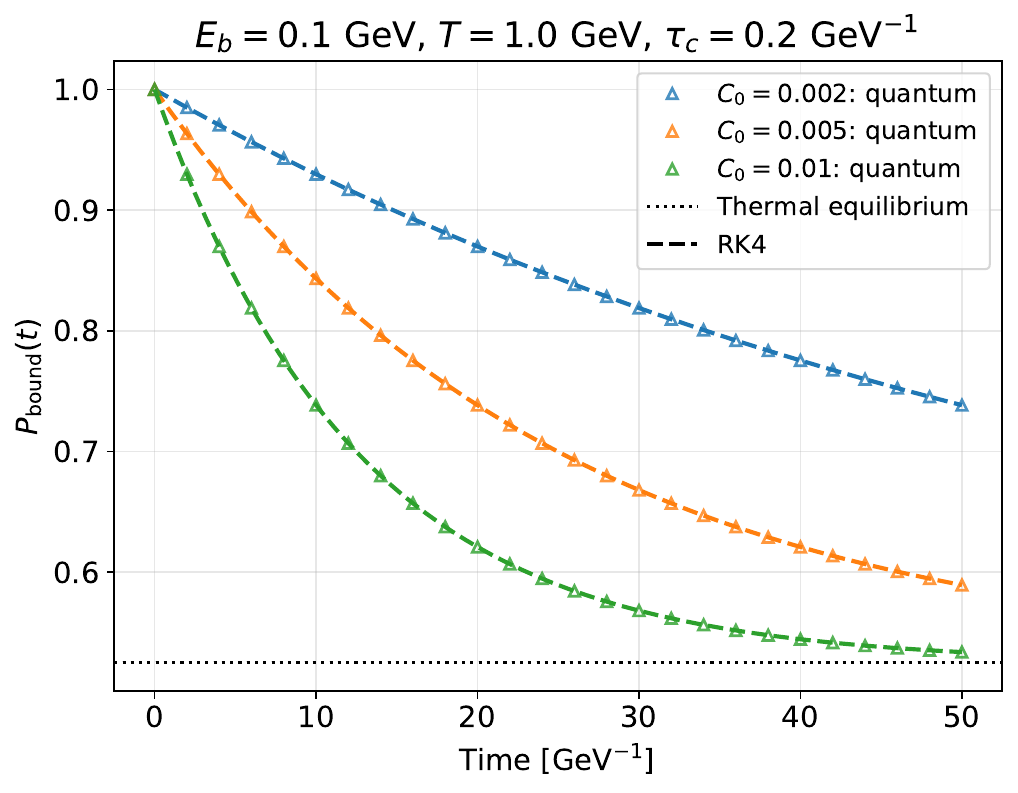}
\caption{Survival probability $P_{\rm bound}(t)$ of the bound state in the Markovian approximation [i.e., Eq.~\eqref{eq:markovian}] for three values of the bath coupling $C_0$ shown in the legend, at fixed $\Delta E\equiv E_b=0.1$ GeV, $T=1.0$ GeV, and bath correlation time $\tau_c=0.2$ GeV$^{-1}$. Dashed curves show the RK4 solution, while the representative triangular markers show the quantum-circuit result. The simulation parameters are $\Delta t=0.05$ GeV$^{-1}$, $N_{\rm step}=1000$, for a total simulated time of $50$ GeV$^{-1}$.} 
\label{fig:markov}
\end{figure}

In Fig.~\ref{fig:markov}, we show the survival probability $P_{\rm bound}(t)$ of the bound state as a function of time, for three values of the bath coupling $C_0 = 0.002$, $0.005$, and $0.01$~GeV$^4$, at a constant $\Delta E = 0.1$~GeV, medium temperature $T = 1.0$~GeV, and the correlation time $\tau_c = 0.2$~GeV$^{-1}$. (See the colored triangular markers.) Note that, as mentioned for the Markovian approximation, these values satisfy $E_b \tau_c\ll 1$. The corresponding dissociation and recombination rates are the ones mentioned earlier below Eq.~\eqref{eq:gamma} and are implemented by replacing $\omega=\Delta E$. Without loss of generality, we take $d_{bf}=1$ GeV$^{-1}$ and initialize the simulation with the pure bound-state density matrix, $\rho_S(0) = |b\rangle\langle b|$.   The dashed curves are obtained by numerically solving the Lindblad equation given in Eq.~\eqref{eq:lindblad} using the fourth-order Runge-Kutta (RK4) method with the same input parameters as mentioned above. The dotted line denotes the equilibrium distribution at the given temperature $T$.

The colored open triangles are obtained from a quantum-circuit implementation of Eq.~\eqref{eq:lindblad} using Qiskit-2.5.1~\cite{Qiskit}. As explained earlier, to perform the simulation, the Lindblad generator is first cast into a completely positive and trace-preserving Kraus map, which is then embedded into a unitary circuit via a Stinespring dilation. The subsystem qubit is paired with a set of ancilla qubits. A single unitary gate acts on both the subsystem and ancillas. At the end of each time step, the ancilla qubits are refreshed for the next iteration. The joint unitary is applied via Qiskit's \texttt{UnitaryGate}, and the ancillas are traced out (\texttt{partial\_trace}) to recover the updated subsystem density matrix. This process is repeated for $N_{\rm step}$ steps to build up the full trajectory $P_{\rm bound}(t)$ as shown in Fig.~\ref{fig:markov}. The simulation parameters are $\Delta t = 0.05$~GeV$^{-1}$ and $N_{\rm step} = 1000$, yielding a total simulated time of $50$~GeV$^{-1}$. We clearly observe that the quantum-circuit result is in great agreement with the numerical RK4 method.

\subsection{Non-Markovian dynamics}
\label{sec:nonmarkovian}

\subsubsection{Pseudomode formulation}
We now attempt to solve Eq.~\eqref{eq:master}  without imposing either of the two approximations that led to Eq.~\eqref{eq:lindblad}, i.e., we neither replace $\rho_S(s)$ by $\rho_S(t)$ inside the memory integral, nor extend the upper limit of integration to infinity. 
By contrast, we keep considering the same exponential correlator with isotropic medium as used in the Markovian case, $C_{ij}(t-s)=\delta_{ij}C_0e^{-|t-s|/\tau_c}$ for illustration.
Since the dipole operator does not depend on any hierarchy between $\tau_E$ and $\tau_S$, it remains the same as in the previous case. This is also because the dipole operator evolves under $H_S$ alone. Simplifying  Eq.~\eqref{eq:master} with the above-mentioned assumptions, we find
\begin{align}
\frac{d\rho_S(t)}{dt}& = -i[H_S,\rho_S(t)]
- |d_{bf}|^2\int_0^t ds\;C_0\,e^{-(t-s)/\tau_c}\nonumber\\
&\times\left(\big[A(t),A(s)\rho_S(s)\big] + \big[\rho_S(s)A(s),A(t)\big]\right).
\label{eq:nonmarkovian}
\end{align}
where $A(t) \equiv \hat L^\downarrow\,e^{-i\Delta E\,t} + \hat L^\uparrow\,e^{i\Delta E\,t}$. Note that Eq.~\eqref{eq:nonmarkovian} retains the history of $\rho_S$ through $\rho_S(s)$ inside the integral, and it is an integro-differential equation, i.e., $\dot\rho_S(t)$ depends on $\rho_S(s)$ at all earlier times $s\le t$. 
Physically, this is the regime when $\tau_E\sim\tau_S$, or equivalently $ E_b\,\tau_c\gtrsim1$. Since we cannot write Eq.~\eqref{eq:nonmarkovian}  in the Lindblad form, the same construction used in section~\ref{sec:markovian} does not apply directly; this is because the Markovian case relied on $\mathcal O_{\rm Markov}$, which is a fixed, time-independent generator acting on $\rho_S(t)$ alone. 

To implement the non-Markovian evolution on a quantum circuit, we enlarge the subsystem by introducing an auxiliary degree of freedom, the pseudomode $P$, which serves as an explicit carrier of the environmental memory. The basic idea is to replace the original structured environment, whose finite correlation time makes the reduced subsystem dynamics nonlocal in time, by a larger system consisting of the subsystem and the pseudomode coupled to a residual Markovian bath. The subsystem can exchange excitations coherently with the pseudomode, so that information transferred from the subsystem is temporarily stored in $P$ and can subsequently flow back, thereby generating memory effects. At the same time, the pseudomode relaxes into the residual Markovian bath, causing this stored information to decay over a finite timescale. We choose the pseudomode relaxation dynamics such that its two-point correlation function has the same exponential time dependence as the original medium correlator $C(\tau)$ as shown in Eq.~\eqref{eq:mcorr}. As a result, tracing out the pseudomode reproduces the desired non-Markovian memory kernel for the subsystem, while the enlarged subsystem-pseudomode state obeys a time-local evolution equation that can be implemented step by step on a quantum circuit.

Having specified the overall pseudomode strategy, we now choose its dynamics to reproduce the particular kernel used in this work.
The target memory kernel $C(\tau)$ is real and contains no oscillatory phase. A nonzero pseudomode energy $H_P=\omega_P |1\rangle \langle 1|$ would introduce an oscillatory factor $e^{-i\omega_P t}$ in its two-point correlation function; We therefore take 
\begin{equation}
    H_P=0, \label{eq:HP}
\end{equation}
leaving only the exponential decay required to reproduce the temporal dependence of the target kernel.\footnote{In a realistic scenario, the medium correlator is computed within the hard thermal loop perturbation theory where one may require $H_P\neq0$.} We couple this pseudomode to the subsystem through Hamiltonian $H_{\rm int}$ which incorporates the same exchange structure as that of the dipole interaction.  Note that this makes the pseudomode a two-level system with its two levels representing excitation and de-excitation. The excitation of the pseudomode is driven by the thermal medium, which is absent in a cold bath.  The key requirement is that $H_{\rm int}$ can only exchange a quantum of excitation between the subsystem and $P$ without independently creating or destroying one. This means whenever the subsystem absorbs a quantum, i.e., $|f\rangle \to |b\rangle$, the pseudomode must lose one, i.e., $|1\rangle \to |0\rangle$, or vice versa. Therefore, $H_{\rm int}$ contains terms like $\hat L^\downarrow$ paired with $\hat L_P^\uparrow$, with $\hat L_P^\uparrow=|0\rangle \langle 1|$ being a jump operator of the pseudomode. Following this, the interaction Hamiltonian between the pseudomode and the subsystem can be written as
\begin{equation}
H_{\rm int} = \lambda_{\rm pm}\big(\hat L^\downarrow\otimes\hat L_P^\uparrow + \hat L^\uparrow\otimes\hat L_P^\downarrow\big),
\label{eq:hint}
\end{equation}
where $\lambda_{\rm pm}$ is the coupling strength between the subsystem and the pseudomode and is fixed by performing a matching to reproduce the same temporal structure as in Eq.~\eqref{eq:mcorr}. 
To this end, we follow the steps given in section~\ref{sec:setup} and write the total Hamiltonian of the system as
\begin{equation}
    H=H_S+H_P+H_{\rm int} = H_S + H_{\rm int},
\end{equation}
where we use the assumption in Eq.~\eqref{eq:HP} in the second equality. This yields the unitary matrix
\begin{equation}
U_I(t) = \mathcal{T}\exp\left[-i\int_0^t dt_1\, H_{{\rm int},I}(t_1)\right]    
\end{equation}
where operators are dressed as $O_I(t)=U_0^\dagger(t)\,O\,U_0(t)$ with $U_0(t)=e^{-iH_St}$. The total density matrix (subsystem+pseudomode) can be evolved as
\begin{equation}
\rho^{\rm PS}_I(t) = U_I(t)\,\rho^{\rm PS}_I(0)\,U_I^\dagger(t), 
\end{equation}

\subsubsection{Pseudomode correlation functions and parameter matching}
For the purpose of matching the pseudomode construction to the original memory kernel in $\lambda_{\rm pm}$, we employ the same type factorization used in section~\ref{sec:setup}: $\rho_{SP}(t)\simeq\rho_S(t)\otimes\rho_P$ with $\rho_P=p_-|0\rangle\langle0|+p_+|1\rangle\langle1|$ where $p_{+}/p_{-}$ are pseudomode populations in the excited/ground state. 
This factorization is used only in deriving the effective reduced equation; the subsequent joint evolution of $S+P$ which will appear in Eq.~\eqref{eq:pseudomode} retains the dynamically generated subsystem-pseudomode correlations.
Repeating the mathematical steps to reach Eq.~\eqref{eq:rhoIdt} and tracing out the pseudomode, we find
\begin{align}
\frac{d\rho_S(t)}{dt} = -\int_0^t ds\;\mathrm{Tr}_P\Big[H_{{\rm int},I}(t),\big[H_{{\rm int},I}(s),\rho_S(s)\otimes\rho_P\big]\Big],
\label{eq:pm_master}
\end{align}
where we utilize the relation $\mathrm{Tr}_P[H_{{\rm int},I}(t)\rho_P]=0$ and the interaction Hamiltonian now reads as 
\begin{equation}
    H_{{\rm int},I}(t)=\lambda_{\rm pm}\left(e^{-i\Delta E t}\,\hat L^\downarrow\otimes\hat L_P^\uparrow \;+\; e^{i\Delta E t}\,\hat L^\uparrow\otimes\hat L_P^\downarrow\right).
\end{equation}
Expanding the product $H_{{\rm int},I}(t)H_{{\rm int},I}(s)$, we obtain
\begin{align}
&H_{{\rm int},I}(t)H_{{\rm int},I}(s) = \lambda_{\rm pm}^2\left(e^{-i\Delta E(t-s)}\,\hat L^\downarrow\hat L^\uparrow\otimes\hat L_P^\uparrow(t)\hat L_P^\downarrow(s)\right.\nonumber\\
&\left.\qquad\quad+ e^{i\Delta E(t-s)}\,\hat L^\uparrow\hat L^\downarrow\otimes\hat L_P^\downarrow(t)\hat L_P^\uparrow(s)\right), 
\label{eq:hint_product}
\end{align}
where other terms such as the ones proportional to $\hat L^\downarrow\hat L^\downarrow$ and $\hat L^\uparrow\hat L^\uparrow$ vanish. 
Now substituting Eq.~\eqref{eq:hint_product} into Eq.~\eqref{eq:pm_master} and tracing over $P$, we finally obtain the time evolution equation of the density matrix:
\begin{widetext}
\begin{equation}
\frac{d\rho_S}{dt} = -\lambda_{\rm pm}^2\int_0^t ds\;\Big\{e^{-i\Delta E(t-s)}\,\mathcal C_-(t-s)\,\big[\hat L^\downarrow,\hat L^\uparrow\rho_S(s)\big] + e^{i\Delta E(t-s)}\,\mathcal C_+(t-s)\,\big[\hat L^\uparrow,\hat L^\downarrow\rho_S(s)\big] + \text{H.c.}\Big\}.
\label{eq:pm_evolution}  
\end{equation}    
\end{widetext}
where H.c. stands for the usual hermitian conjugate and $\mathcal{C}_{\pm}(\tau)$ are pseudomode correlators:
\begin{align}
&\mathcal C_-(\tau) \equiv \mathrm{Tr}_P\big[\hat L_P^\uparrow(\tau)\hat L_P^\downarrow(0)\,\rho_P\big],\nonumber\\ 
&\mathcal C_+(\tau) \equiv \mathrm{Tr}_P\big[\hat L_P^\downarrow(\tau)\hat L_P^\uparrow(0)\,\rho_P\big],
\label{eq:pcorr}
\end{align}
Equation~\eqref{eq:pm_evolution} is the exact analog of Eq.~\eqref{eq:master}, with the pseudomode replacing the medium correlator in Eq.~\eqref{eq:mcorr}. To determine $\lambda_{\rm pm}$, 
we therefore need to evaluate the pseudomode correlation functions $\mathcal C_\pm(\tau)$, which follow from the pseudomode dynamics generated by its coupling to the residual bath.

The coupling of the pseudomode to the rest of the Markovian plasma is  described by a Lindblad dissipator of the form
\begin{equation}
\mathcal{L}_P[\rho_{SP}] = \kappa_{\rm pm}(1+\bar{n}_{\rm pm})\, \mathcal{D}[\hat{L}_P^{\uparrow}]\rho_{SP} + \kappa_{\rm pm}\bar{n}_{\rm pm}\, \mathcal{D}[\hat{L}_P^{\downarrow}]\rho_{SP},
\label{eq:pseudomode_dissipator}
\end{equation}
where $\kappa_{\rm pm}$ is related to the pseudomode relaxation rate, which we fix by matching the pseudomode relaxation rate with Eq.~\eqref{eq:mcorr}.
Equation~\eqref{eq:pseudomode_dissipator} is the pseudomode analog of the dissipative terms in Eq.~\eqref{eq:markovian}; the apparent reversal between the arrow directions and the thermal occupation factors arises because excitation gained by the subsystem corresponds to de-excitation of the pseudomode, and vice versa.
Since $\mathcal{L}_P$ is a Lindblad generator, the reduced pseudomode density matrix also relaxes according to 
\begin{equation}
    \frac{d\rho_P}{dt}=-i[H_P,\rho_P]+\mathcal L_P[\rho_P]=\mathcal L_P[\rho_P], \label{eq:rhopevolve}
\end{equation}
where we use $H_P=0$.
Plugging $p_+$ into Eq.~\eqref{eq:rhopevolve}, we find that the population $p_+$ obeys 
\begin{equation}
    \dot p_+ = \kappa_{\rm pm}\bar n_{\rm pm}\,p_- - \kappa_{\rm pm}(1+\bar n_{\rm pm})\,p_+.
\end{equation}
For the purpose of evaluating the stationary pseudomode correlation functions, we take \(\rho_P\) to be the stationary state of the thermal dissipator \(\mathcal L_P\), i.e., $\dot p_+=0$.
Therefore, using $\dot p_+=0$ together with $p_-=1-p_+$, we obtain 
\begin{equation}
p_+ = \frac{\bar n_{\rm pm}}{1+2\bar n_{\rm pm}}, \qquad p_- = \frac{1+\bar n_{\rm pm}}{1+2\bar n_{\rm pm}}.
\label{eq:p0p1}
\end{equation}

We next evaluate the correlators given in Eq.~\eqref{eq:pcorr}. To this end, we need to compute how the pseudomode coherence decays with time because the correlators $\mathcal C_\pm(\tau)=\langle \hat L_P^{\uparrow(\downarrow)}(\tau)\hat L_P^{\downarrow(\uparrow)}(0)\rangle$ are constructed by first acting with the operator $L_P^{\downarrow(\uparrow)}$ on the pseudomode's steady state and then letting the resulting state evolve for a time $\tau$ before the second operator is applied. Since $\rho_P$ is initially diagonal, acting with either $\hat L_P^\downarrow$ or $\hat L_P^\uparrow$ converts this to an off-diagonal (coherence) term, i.e., $\hat L_P^\downarrow\rho_P\propto\ket1\bra0$ and $\hat L_P^\uparrow\rho_P\propto\ket0\bra1$.  
Again, the time evolution of the pseudomode is governed by the dissipator $\mathcal L_P$. 
Therefore, e.g., for $\rho_P^{10}=|1\rangle \langle0|$, we have
\begin{equation}
\frac{d\rho_{P}^{10}}{dt}=\mathcal{L}_P[\rho_P^{10}]=-\Gamma_c\,\rho_{P}^{10},   
\label{eq:pmevolve}
\end{equation}
where pseudomode correlator decay rate $\Gamma_c$ is defined as
\begin{equation}
    \Gamma_c = \frac{\kappa_{\rm pm}}{2}(1+2\bar n_{\rm pm}).
\end{equation}
This describes the time evolution of the density matrix, leading to the solution
\begin{equation}
    \rho_P^{10}(t) = \rho_P^{10}(0)e^{-\Gamma_c t}, \label{eq:timeevolverhocoherence}
\end{equation}
and the same expression goes through for $\rho_P^{01}$ as well. 

We are now ready to calculate the pseudomode correlators in Eq.~\eqref{eq:pcorr}. Beginning with $\rho_P(0)=p_+|1\rangle \langle 1|+p_-|0\rangle \langle 0|$, $\mathcal{C}_-(\tau)$ is given by
\begin{equation}
    \mathcal{C}_-(\tau)={\rm Tr}\Big[\hat{L}_P^\uparrow(\tau)\left.\left(p_-|1\rangle\langle 0|\right)\right|_{t=0} \Big].
\end{equation}
Since operator $\hat{L}_P^\uparrow$ acts on the state at $t=\tau$, the $p_-|1\rangle\langle0|$ evaluated at $t=0$ should be evolved to $t=\tau$ using Eq.~\eqref{eq:timeevolverhocoherence}. We thus obtain
\begin{equation}
    \mathcal C_-(\tau) =  \mathrm{Tr}\Big[\hat L_P^\uparrow(\tau)\cdot p_-e^{-\Gamma_c\tau}|1\rangle\langle0|\Big]=  p_-\,e^{-\Gamma_c\tau}. \label{eq:corval1}
\end{equation}
In the same manner, we further find
\begin{equation}
\mathcal C_+(\tau) =  \mathrm{Tr}\Big[\hat L_P^\downarrow(\tau)\cdot p_+e^{-\Gamma_c\tau}|0\rangle\langle1|\Big]
=  p_+\,e^{-\Gamma_c\tau}.
\label{eq:corval2}
\end{equation}
As a consistency check, in the limit $\bar{n}_{\rm pm}\to 0$, $p_+=0$ and $p_-=1$, so that $\mathcal C_+$ vanishes and only pseudomode de-excitation contributes. This is expected because, in a cold bath, the pseudomode cannot absorb excitations from the environment, whereas such excitation processes are allowed at finite temperature. 

We finally perform the matching in the zero-temperature limit because the target kernel in Eq.~\eqref{eq:mcorr} is a single real exponential and does not distinguish thermal excitation and de-excitation channels. At finite temperature, the pseudomode develops two distinct correlators, \(C_\pm\), and therefore represents a thermal extension of the bare kernel rather than an exact one-to-one reproduction of Eq.~\eqref{eq:mcorr}.
Comparing the correlator in Eq.~\eqref{eq:corval1} at $\bar n_{\rm pm}\to 0$ with the actual memory kernel in Eq.~\eqref{eq:mcorr}, we demand $\Gamma_c=\tau_c^{-1}$, hence $\kappa_{\rm pm}=2/\tau_c$. 
Having fixed the decay profile, we then match the overall strength of the surviving kernel $\mathcal{C}_-(\tau)$ in Eq.~\eqref{eq:pm_evolution} to that of the original kernel in Eq.~\eqref{eq:nonmarkovian}, yielding
\begin{equation}
    \lambda_{\rm pm}^2=|d_{bf}|^2C_0.
\end{equation}

\subsubsection{Quantum-circuit implementation}
Having fixed the pseudomode parameters by matching its correlation functions to the target memory kernel, we now evolve the full subsystem-pseudomode density matrix according to
\begin{equation}
\frac{d\rho_{SP}}{dt} = -i[H_S + H_{\rm int},\,\rho_{SP}] + \mathcal L_P[\rho_{SP}],
\label{eq:pseudomode}
\end{equation}
where \(\rho_{SP}\) denotes the full joint density matrix of the subsystem and pseudomode. In contrast to the factorization employed above for the perturbative kernel matching, the evolution in Eq.~\eqref{eq:pseudomode} retains the subsystem-pseudomode correlations dynamically generated by \(H_{\rm int}\), while $\rho_{SP}(0)=\rho_S(0)\otimes\rho_P(0)$. The subsystem density matrix is obtained by tracing out the pseudomode, i.e., $\rho_S(t) = \mathrm{Tr}_P[\rho_{SP}(t)]$. 
It is worth mentioning that while the zero-temperature evolutions of Eq.~\eqref{eq:pseudomode} and Eq.~\eqref{eq:nonmarkovian} are the same, the finite-temperature counterpart is different. This is because Eq.~\eqref{eq:nonmarkovian} does not contain finite-temperature memory effects while we self-consistently incorporate them in pseudomode construction.  

To implement the non-Markovian dynamics, i.e., Eq.~\eqref{eq:pseudomode}, on a quantum circuit, we use Lie-Trotter splitting per interval $\Delta t$. Note that unlike the Markovian case in Eq.~\eqref{eq:lindblad}, the non-Markovian generator naturally decomposes into two non-commuting generators. First, the coherent term containing the Hamiltonian $H_S+H_{\rm int}$, which couples the subsystem with the pseudomode, and second, the dissipative term $\mathcal L_P$ which acts on the pseudomode.  We approximate the evolution over one time step by applying them one after the other.  A simple ordering such as coherent followed by dissipative carries an error of order $\Delta t^2$ at each time step. To improve this, we use the symmetric ordering as represented in Fig.~\ref{fig:nonmarkov}
\begin{equation}
e^{\mathcal L_{SP}\Delta t}[\rho_{SP}] \approx U_{SP}\Big(e^{\mathcal L_P\Delta t}\big[U_{SP}\,\rho_{SP}\,U_{SP}^\dagger\big]\Big)U_{SP}^\dagger,  
\label{eq:nonmarkovstep}
\end{equation}
where $e^{\mathcal L_{SP}\Delta t}[\rho_{SP}]$ denotes the state $\rho_{SP}(t+\Delta t)$ obtained by evolving $\rho_{SP}(t)$ under the full generator for time $\Delta t$. $\mathcal{L}_{SP}$ is the generator in Eq.~\eqref{eq:pseudomode} and 
\begin{equation} 
    U_{SP}\equiv e^{-i(H_S+H_{\rm int})\Delta t/2},\label{eq:Usp}
\end{equation}
which splits the coherent evolution in two halves around the dissipative step. This split of the ordering reduces the error at each time step to $\Delta t^3$. The overall error after accumulating many repetitions is of the order of $\Delta t^2$. Note that for our specific case, the Hamiltonian in $U_{SP}$ has a very simple form of coupling the pseudomode and subsystem, which allows the full Hamiltonian $H_S+H_{\rm int}$ to be solved exactly.

\begin{figure}[h]
\centering 
\includegraphics[width=1\linewidth]{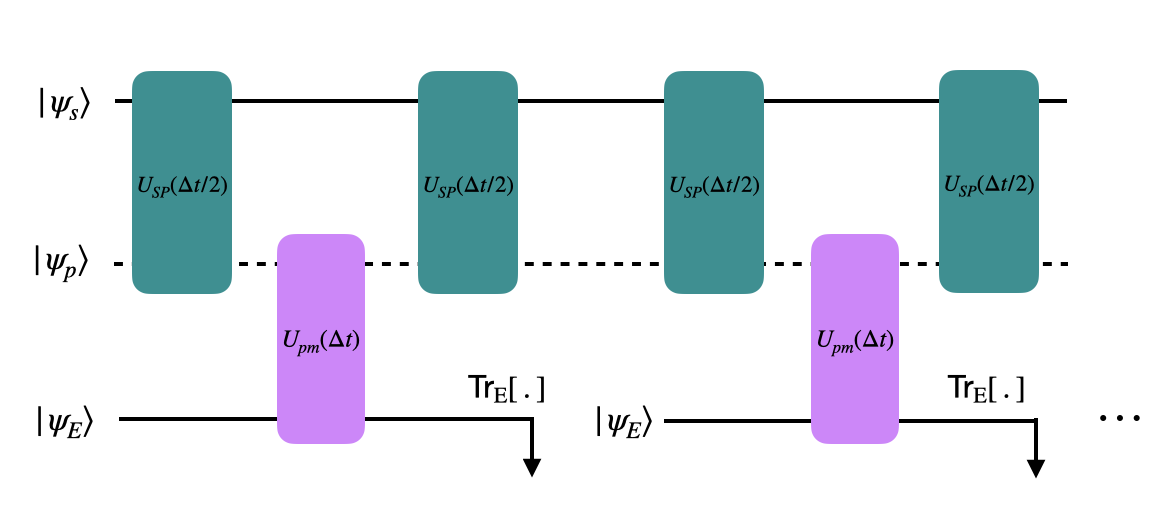}
\caption{Circuit realization of two non-Markovian time steps. The subsystem qubit $|\psi_S\rangle$ and pseudomode qubit $|\psi_P\rangle$ (dashed line) both persist across repetitions and are never reset. Each step applies the symmetric splitting: a coherent half-step $U_{SP}(\Delta t/2)$ acting on the subsystem and pseudomode; the dissipative step $U_{\rm pm}(\Delta t)$ acting on the pseudomode and a fresh environment ancilla $|\psi_E\rangle$, which is traced out immediately afterward; and a second coherent half-step $U_{SP} (\Delta t/2)$ on the subsystem and pseudomode. Repeating this three-unitary sequence $N_{\rm step}$ times describes the  evolution over total time $t=N_{\rm step}\Delta t$.} 
\label{fig:nonmarkov}
\end{figure} 

The dissipative step of the pseudomode is implemented in a similar way as mentioned earlier for the Markovian case in section~\ref{sec:markovian} with decay and excitation rates $\kappa_{\rm pm}(1+\bar n_{\rm pm})$ and $\kappa_{\rm pm}\bar n_{\rm pm}$ plugged in as jump rates. As mentioned earlier, we set the pseudomode Hamiltonian $H_P=0$. With the generator of pseudomode relaxation, we exponentiate the evolution over a given time step $\Delta t$.  Similar to the previous case, from this evolution we obtain Kraus operators.  For our case, we get four nonzero Kraus operators that are reduced to two in the limit $\bar n_{\rm pm}=0$.  Following the same procedure as in section~\ref{sec:markovian}, the Kraus operators are then used to construct the unitary operator $U_{\rm pm}$ which acts on both the pseudomode and a fresh ancilla. Throughout this entire dissipation step, the subsystem qubit remains the same.

Evolving the system by one time step $\Delta t$ requires three separate unitary operations which are applied one after another in a fixed order. The first operation is the half-step $U_{SP}(\Delta t/2)$, i.e., the one defined in Eq.~\eqref{eq:Usp}, which is applied to the subsystem and pseudomode together; the second step is to prepare a fresh ancilla and apply the dissipative step $U_{\rm pm}$ to the pseudomode and ancilla together. At the end of this step, the ancilla is discarded and prepared fresh for the next time step. Finally, another half-step operation $U_{SP}(\Delta t/2)$ is applied, which completes one time step. Repeating this three-part sequence $N_{\rm step}$ times evolves the system through a total simulated time $t=N_{\rm step}\Delta t$. The subsystem and pseudomode qubits are carried over from one repetition to the next one. On the other hand, a fresh ancilla is prepared at the start of every repetition and discarded at the end of it.  The circuit needs one qubit for the subsystem, one qubit for the pseudomode, and enough ancilla qubits to hold Kraus operators, i.e., $n_{\rm anc}=\lceil\log_2(n_{\rm ops})\rceil$ where $n_{\rm ops}$ is the number of Kraus operators. 

\begin{figure}[h]
\centering 
\includegraphics[width=1\linewidth]{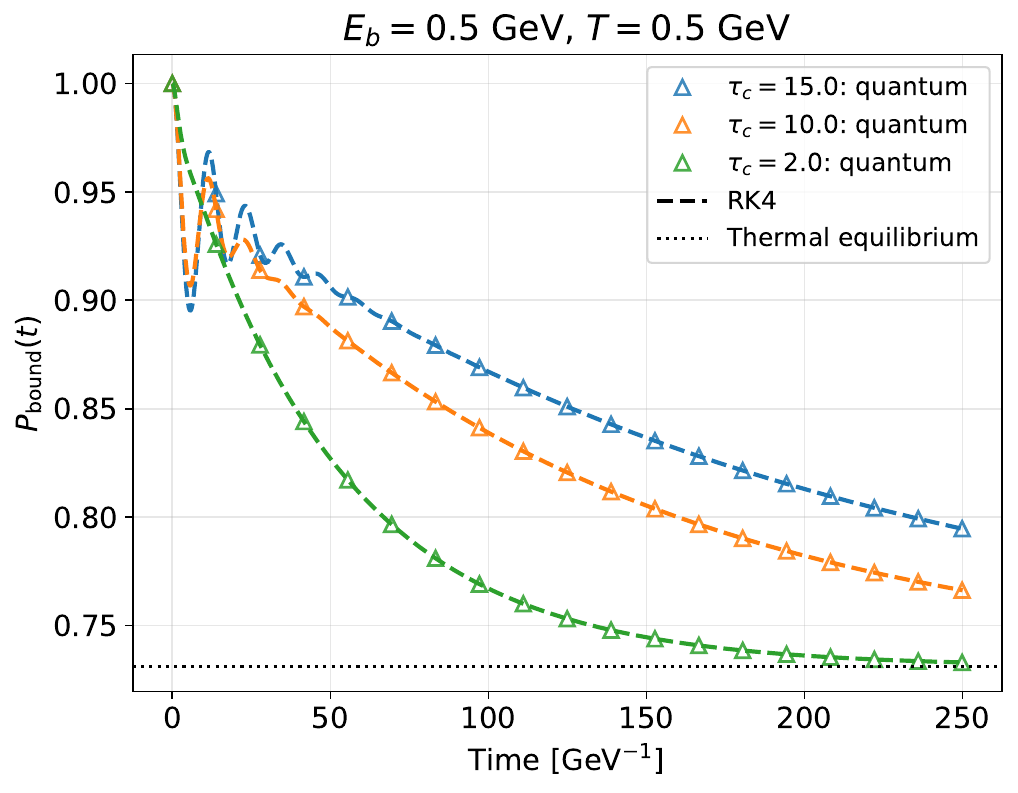}
\caption{Survival probability $P_{\rm bound}(t)$ of the bound state in non-Markovian case [i.e., Eq.~\eqref{eq:pseudomode}] for three values of bath memory time $\tau_c$, at fixed $\Delta E\equiv E_b=0.5$ GeV, $T=0.5$ GeV, and coupling $C_0=0.01$ GeV. Dashed curves show the RK4 solution, while the triangular markers show the quantum-circuit result using pseudomode. The simulation parameters are $\Delta t=0.05$ GeV$^{-1}$ and $N_{\rm step}=5000$, resulting in a total simulated time of $250$ GeV$^{-1}$.} 
\label{fig:nonmarkov}
\end{figure}

In Fig.~\ref{fig:nonmarkov}, we show the survival probability $P_{\rm bound}(t)$ of the bound state as a function of time, for three different values of the bath memory time $\tau_c = 15.0$, $10.0$, and $2.0$~GeV$^{-1}$ and at a constant value of $\Delta E = 0.5$~GeV, $T = 0.5$~GeV, and bath coupling $C_0 = 0.01$~GeV$^2$. For these parameters, $E_b \tau_c \gtrsim 1$, placing the system outside the regime where the validity of the Markovian approximation is guaranteed. Therefore, we explicitly keep the pseudomode and solve the full non-Markovian evolution equation for the joint system-pseudomode density matrix $\rho_{SP}(t)$, Eq.~\eqref{eq:pseudomode}. We initialize the simulation with the subsystem in the pure bound state and the pseudomode in its ground state, i.e., $\rho_{SP}(0) = |{b},0\rangle\langle {b},0|$. The dashed curves are obtained by numerically integrating the non-Markovian evolution equation again using the RK4 method applied to the full joint generator acting on the $4\times4$ system-pseudomode density matrix, with no splitting of the coherent and dissipative parts.

The open triangles shown in Fig.~\ref{fig:nonmarkov} are obtained from the quantum-circuit implementation described above. Following the scheme as discussed in Eq.~\eqref{eq:nonmarkovstep}, at each time step, the coherent subsystem and pseudomode evolution under $H_S + H_{\rm int}$ is applied as two half-steps, which are symmetrically separated around the dissipative step acting on the pseudomode and a freshly prepared ancilla register. For the dissipative step, the Lindblad generator $\mathcal{L}_P$ is exponentiated for a given time interval $\Delta t$ to obtain $M(\Delta t) = \exp(\mathcal{L}_P \Delta t)$, which is then used to extract the set of Kraus operators as described above for the Markovian case.  The Kraus operators are then used to construct the unitary matrix $U_{\rm pm}$. On the quantum circuit, $U_{\rm pm}$ is implemented as a single unitary gate acting on pseudomode and ancilla qubits. After the gate is applied, the ancilla qubits are traced out and initialized fresh at the start of each time step.  The coherent half-steps $U_{SP}(\Delta t/2)$ are likewise implemented as a two-qubit unitary gate obtained from $H_S + H_{\rm int}$. To keep the memory effects, the subsystem and pseudomode qubits remain unchanged across repetitions while the ancilla register is discarded and re-prepared at the start of every step.  It is worth mentioning that at the end of each step, $P_{\rm bound}(t)$ is recorded by tracing out the pseudomode from the joint state. This trace is performed only for observable extraction and does not affect the evolving joint states. The simulation parameters are $\Delta t = 0.05$~GeV$^{-1}$ and $N_{\rm step} = 5000$, yielding a total simulated time of $250$~GeV$^{-1}$. The quantum-circuit result agrees with the RK4 solution. 

\subsubsection{Markovian limit of the pseudomode construction}
\begin{figure}[h]
\centering 
\includegraphics[width=1\linewidth]{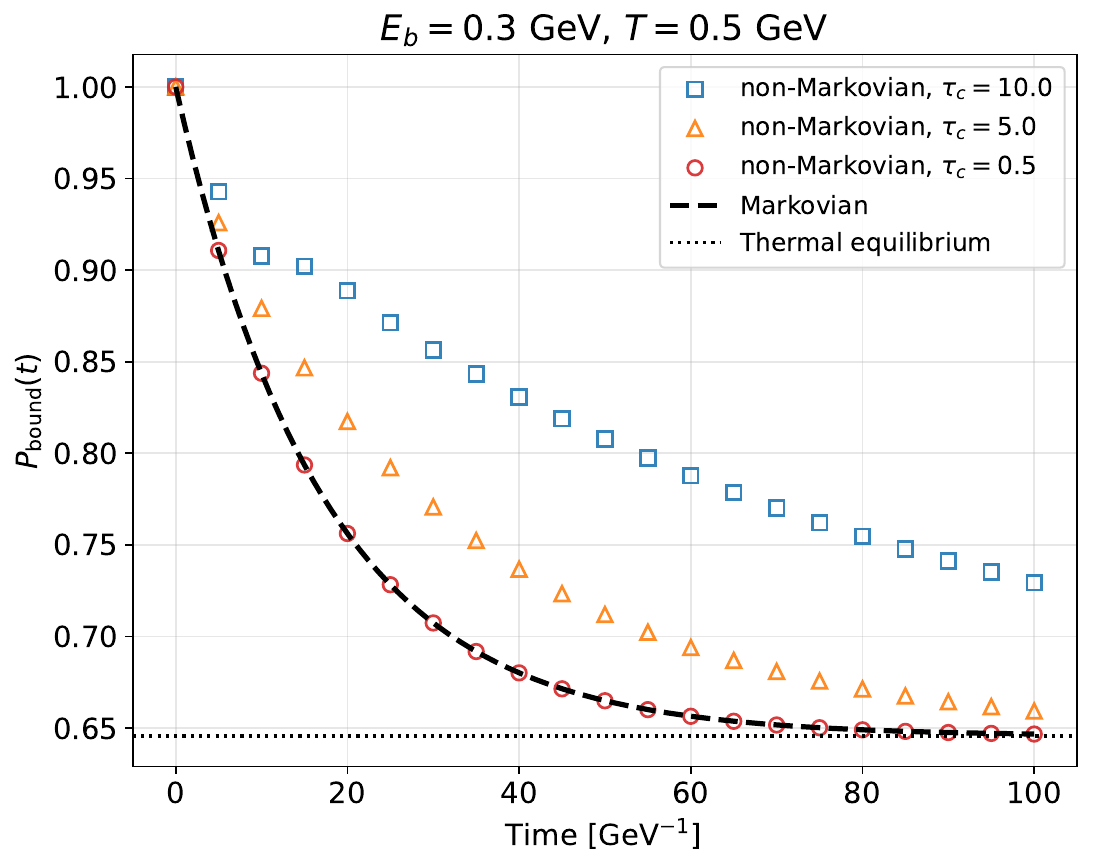}
\caption{Survival probability $P_{\rm bound}(t)$ of the bound state as a function of time at  $\Delta E\equiv E_b = 0.3$~GeV, $T = 0.5$~GeV, and $\lambda_{\rm pm}^2 = 0.5$~GeV. The dashed curve shows the Markovian solution [i.e., Eq.~\eqref{eq:lindblad_corrected}], while the square, triangle, and open-circle markers display the non-Markovian [i.e., Eq.~\eqref{eq:pseudomode}] quantum-circuit-based computations with $\tau_c = 10.0$, $5.0$, and $0.5$~GeV$^{-1}$, respectively. As $\tau_c$ decreases, the non-Markovian curves converge onto the Markovian reference calculated with $\tau_c=0.1~{\rm GeV}^{-1}$. The simulation parameters are $\Delta t = 0.01$~GeV$^{-1}$, $N_{\rm step} = 10000$, resulting in a total simulated time of $100$~GeV$^{-1}$.} 
\label{fig:comparison}
\end{figure}

We are now in the position to demonstrate the Markovian limit of the non-Markovian evolution performed using the joint evolution of the pseudomode and subsystem (see Fig.~\ref{fig:comparison}). Since $\tau_c$ controls bath memory time in the correlator, one naive expectation is that the non-Markovian evolution should approach the Markovian one with decreasing bath correlation time for a given $E_b$. To this end, we keep $E_b$ fixed and vary $\tau_c$.  As a self-consistency check, we take the decay profile of the pseudomode, i.e., $p_{\pm}\,e^{-\Gamma_c \tau}$, and take its Fourier transform to define
\begin{equation}
\tilde{\gamma}_\pm(\omega) = \frac{2\, p_{\pm} \Gamma_c}{\Gamma_c^2+\omega^2},
\label{eq:gamma1}
\end{equation}
where $\Gamma_c$ is the decay rate of the pseudomode and $p_{\pm}$ are same as those shown in Eq.~\eqref{eq:p0p1}. As discussed earlier, since the pseudomode couples to the subsystem through operators $\hat L_P^{\downarrow}$ and $\hat L_P^{\uparrow},$ it is $\Gamma_c$ that sets the effective memory experienced by the subsystem. For consistency, we compute $\Gamma_c$ at a fixed value of $\kappa_{\rm pm}=2/\tau_c$. Note that in the limit $\bar{n}_{\rm pm}=0$, $\Gamma_c$ is the same as the inverse memory time in Eq.~\eqref{eq:mcorr}.  Moreover,  with thermal effects, the pseudomode $\tilde{\gamma}_\pm(\omega)$ in Eq.~\eqref{eq:gamma1} is different from the bare memory kernel given in Eq.~\eqref{eq:gamma}. This is because the pseudomode construction is modeled by incorporating thermal distributions, through both $p_{\pm}$ and $\Gamma_c$, while Eq.~\eqref{eq:gamma} does not contain any thermal term and uses the bare rate $1/\tau_c$ in place of $\Gamma_c$.

For consistency of the Markovian limit of pseudomode construction, we consider the fast-relaxation regime in which the pseudomode no longer needs to be retained as an explicit dynamical degree of freedom. In this limit, the pseudomode can be eliminated, leaving an effective Lindblad equation for the subsystem alone, with transition rates determined by the pseudomode correlation functions. We therefore write
\begin{equation}
\frac{d\rho_S}{dt} = -i[H_S,\rho_S] + \Gamma_{\rm diss}\,\mathcal{D}[\hat L^{\uparrow}]\rho_S + \Gamma_{\rm rec}\,\mathcal{D}[\hat L^{\downarrow}]\rho_S,
\label{eq:lindblad_corrected}
\end{equation}
with the rates evaluated at $\omega=\Delta E$ as before. However, the rates now depend on the decay profile of the pseudomode [see Eq.~\eqref{eq:gamma1}] and are given by
\begin{align}
&\Gamma_{\rm diss} = \lambda_{\rm pm}^2\tilde{\gamma}_-(\Delta E) = \lambda_{\rm pm}^2\frac{2\,p_{-}\Gamma_c}{\Gamma_c^2+\Delta E^2},\nonumber\\
&\Gamma_{\rm rec} = \lambda_{\rm pm}^2\tilde{\gamma}_+(\Delta E) = \lambda_{\rm pm}^2\frac{2\,p_{+}\Gamma_c}{\Gamma_c^2+\Delta E^2}.
\label{eq:corrected_rates}
\end{align}
Note that $\Gamma_{\rm diss}$ and $\Gamma_{\rm rec}$ correspond to decay and excitation of the pseudomode. For a cold bath, the pseudomode cannot absorb any excitations, i.e., $p_+=0$, leading to $\Gamma_{\rm rec}=0$. As a result, only the pseudomode decay gives a finite contribution. On the other hand, at finite $T$ both correlators contribute.  

To obtain the Markovian (memoryless) limit from the pseudomode construction given in Eq.~\eqref{eq:nonmarkovstep}, we use Eq.~\eqref{eq:lindblad_corrected} as the reference to which the pseudomode is expected to converge. Note that Eq.~\eqref{eq:lindblad_corrected} is derived from the pseudomode's decay kernel in the fast relaxation limit. For all values of memory time, we keep bath coupling fixed so that only the effect of memory time $\tau_c$ is probed. In Fig.~\ref{fig:comparison}, we show the survival probability $P_{\rm bound}(t)$ for a constant values $\Delta E=0.3$~GeV, $T=0.5$~GeV, and $\lambda_{\rm pm}^2=0.5$~GeV. The dashed curve shows the Markovian solution from Eq.~\eqref{eq:lindblad_corrected} with $\tau_c$ set to be $0.1$~GeV$^{-1}$; with this choice, the Markovian limit is valid as $E_b \tau_c = 0.03 \ll 1$. The square, triangle, and open-circle markers show the non-Markovian (pseudomode) quantum-circuit-based results at $\tau_c=10.0$, $5.0$, and $0.5$~GeV$^{-1}$, respectively; they are obtained by solving the full joint system-pseudomode evolution described above, with no reduction to Eq.~\eqref{eq:lindblad_corrected}.  As $\tau_c$ decreases, the non-Markovian pseudomode results converge to the effective Markovian evolution characterized by the rates in Eq.~\eqref{eq:corrected_rates}.  Simulation parameters are $\Delta t=0.01$~GeV$^{-1}$ and $N_{\rm step}=10000$, yielding a total simulated time of $100$~GeV$^{-1}$.

\section{Summary and Conclusions}
\label{sec:summary}

In this paper, we study an OQS approach to the evolution of a subsystem density matrix within both the Markovian and non-Markovian regimes using quantum computing methods. As an example, we consider a two-level system such as a heavy particle-antiparticle bound state interacting with an isotropic thermal plasma. To simplify the computations and simulations, we take the bath correlation function to decay exponentially at a rate $1/\tau_c$, with $\tau_c$ being the correlation time of the medium correlator. We first work in the Markovian limit $E_b\,\tau_c\ll1$, where the master equation given in Eq.~\eqref{eq:master} reduces to a time-local Lindblad equation. We convert the corresponding Lindblad generator into a trace-preserving Kraus map and implement it as a unitary circuit through the Stinespring dilation. In Fig.~\ref{fig:markov}, we show that our quantum circuit simulation agrees with the numerical solution obtained using the RK4 method. We also showed that a larger coupling between the bath and the subsystem leads to faster dissociation of the bound state.

We next relax the Markovian approximations to obtain the non-Markovian integro-differential equation governing the evolution of $\rho_S(t)$ given in Eq.~\eqref{eq:nonmarkovian}. In this case, the generator depends on the history of the density matrix rather than its instantaneous value. The non-Markovian dynamics become important when the bath correlation time becomes comparable to or longer than the subsystem's characteristic evolution time, i.e., $ E_b\,\tau_c\gtrsim1$, so that the medium retains memory of the subsystem's past on timescales relevant to the bound-unbound transition. To implement the non-Markovian evolution on a quantum computer, we introduce a two-level auxiliary pseudomode which we couple to the subsystem as well as to a residual Markovian bath through a thermal dissipator. We incorporate the finite temperature effects via thermal distribution functions. We show that in the limit of vanishing thermal occupation, the pseudomode's reduced dynamics reproduces the non-Markovian evolution given in Eq.~\eqref{eq:nonmarkovian} with a memory kernel that decays exponentially at the same rate $1/\tau_c$ as the physical bath correlator. We construct an explicit quantum circuit for the joint evolution of the subsystem and pseudomode using a symmetric splitting between the coherent subsystem-pseudomode coupling and the pseudomode's dissipative relaxation into the thermal bath. This requires one additional qubit compared to the Markovian construction. In Fig.~\ref{fig:nonmarkov}, we show that our quantum circuit results agree with the RK4 method across different values of the bath correlation time. We also show that keeping $E_b$ fixed, a larger correlation time of the medium leads to slower dissociation of the bound state, reflecting the fact that a longer-lived memory suppresses the effective dissipative rate relative to the Markovian case.

Finally, we explicitly show that tracing out the pseudomode recovers the non-Markovian evolution of Eq.~\eqref{eq:nonmarkovian} in the appropriate limit, confirming that the pseudomode captures memory effects in the bath and can be implemented as a time-local circuit. In Fig.~\ref{fig:nonmarkov}, we further show that the non-Markovian construction smoothly converges onto its Markovian limit as $\tau_c\to0$, which validates the consistency of non-Markovian dynamics through the auxiliary pseudomode.

Within the simplified approximations adopted here, our results provide a systematic framework for simulating OQS dynamics in both the Markovian and non-Markovian regimes with quantum computing methods, paving an avenue to simulating the dynamics of hard probes such as heavy quarks, quarkonium, and jets in HICs. The pseudomode approach can be extended to realistic applications.  In the future, a more realistic phenomenological treatment with the medium correlator computed within the hard thermal loop perturbation theory will be explored to extend the applicability of quantum computing to precision calculations of hard probe observables in HICs.

\section*{Acknowledgments}
The work of DK is supported in part by NSF Grant No. PHY-2609913. The work of BS is supported by Fundação para a Ciência e a Tecnologia (FCT) through the ERC-PT A-Projects ‘Unveiling’, financed by PRR, NextGenerationEU and Fundação para a Ciência e a Tecnologia (FCT) under con-
tracts2023.15319.PEX(https://doi.org/10.54499/2023. 15319.PEX).

\bibliography{main}
\end{document}